\documentclass[a4paper,11pt]{article}
\usepackage{pos}

\usepackage{gensymb}
\usepackage[numbers]{natbib} 

\title{EUSO-SPB2 Telescope Optics and Testing}
 \ShortTitle{EUSO-SPB2 Telescope Optics and Testing}

\author*[a]{Viktoria Kungel}
\author[a]{Randy Bachman}
\author[a]{Jerod Brewster}
\author[a]{Madeline Dawes}
\author[a]{Julianna Desiato}
\author[b]{Johannes Eser}
\author[a]{William Finch}
\author[a]{Lindsey Huelett}
\author[b]{Angela V. Olinto}
\author[a]{Justin Pace}
\author[c]{Miroslav Pech}
\author[d]{Patrick Reardon}
\author[c]{Petr Schovanek}
\author[a]{Chantal Wang}
\author[a]{Lawrence Wiencke}

\affiliation[a]{Colorado School of Mines,\\
  Golden, CO, USA}
\affiliation[b]{University of Chicago,\\
Chicago, IL, USA}
\affiliation[c]{Joint Laboratory of Optics,
Palacky University and Institute of Physics AV ČR,\\
Olomouc, Czech Republic}
\affiliation[d]{The University of Alabama in Huntsville,\\
Huntsville, IL, USA}

\forColl{JEM-EUSO} 

\emailAdd{kungel@mines.edu}

\abstract{The Extreme Universe Space Observatory - Super Pressure Balloon (EUSO-SPB2) mission will fly two custom telescopes that feature Schmidt optics to measure \v{C}erenkov- and fluorescence-emission of extensive air-showers from cosmic rays at the PeV and EeV-scale, and search for $\tau$-neutrinos. Both telescopes have 1-meter diameter apertures and UV/UV-visible sensitivity. The \v{C}erenkov telescope uses a bifocal mirror segment alignment, to distinguish between a direct cosmic ray that hits the camera versus the \v{C}erenkov light from outside the telescope. Telescope integration and laboratory calibration will be performed in Colorado. To estimate the point spread function and efficiency of the integrated telescopes, a test beam system that delivers a 1-meter diameter parallel beam of light is being fabricated.
End-to-end tests of the fully integrated instruments will be carried out in a field campaign at dark sites in the Utah desert using cosmic rays, stars, and artificial light sources. Laser tracks have long been used to characterize the performance of fluorescence detectors in the field. For EUSO-SPB2 an improvement in the method that includes a correction for aerosol attenuation is anticipated by using a bi-dynamic Lidar configuration in which both the laser and the telescope are steerable. We plan to conduct these field tests in Fall 2021 and Spring 2022 to accommodate the scheduled launch of EUSO-SPB2 in 2023 from Wanaka, New Zealand.
}

\FullConference{37$^{\rm{th}}$ International Cosmic Ray Conference (ICRC 2021)\\
		July 12th -- 23rd, 2021\\
		Online -- Berlin, Germany}

\begin{document}
\maketitle

\section{Mission}
\begin{figure}[ht]
    \centering
\includegraphics[width=\textwidth]{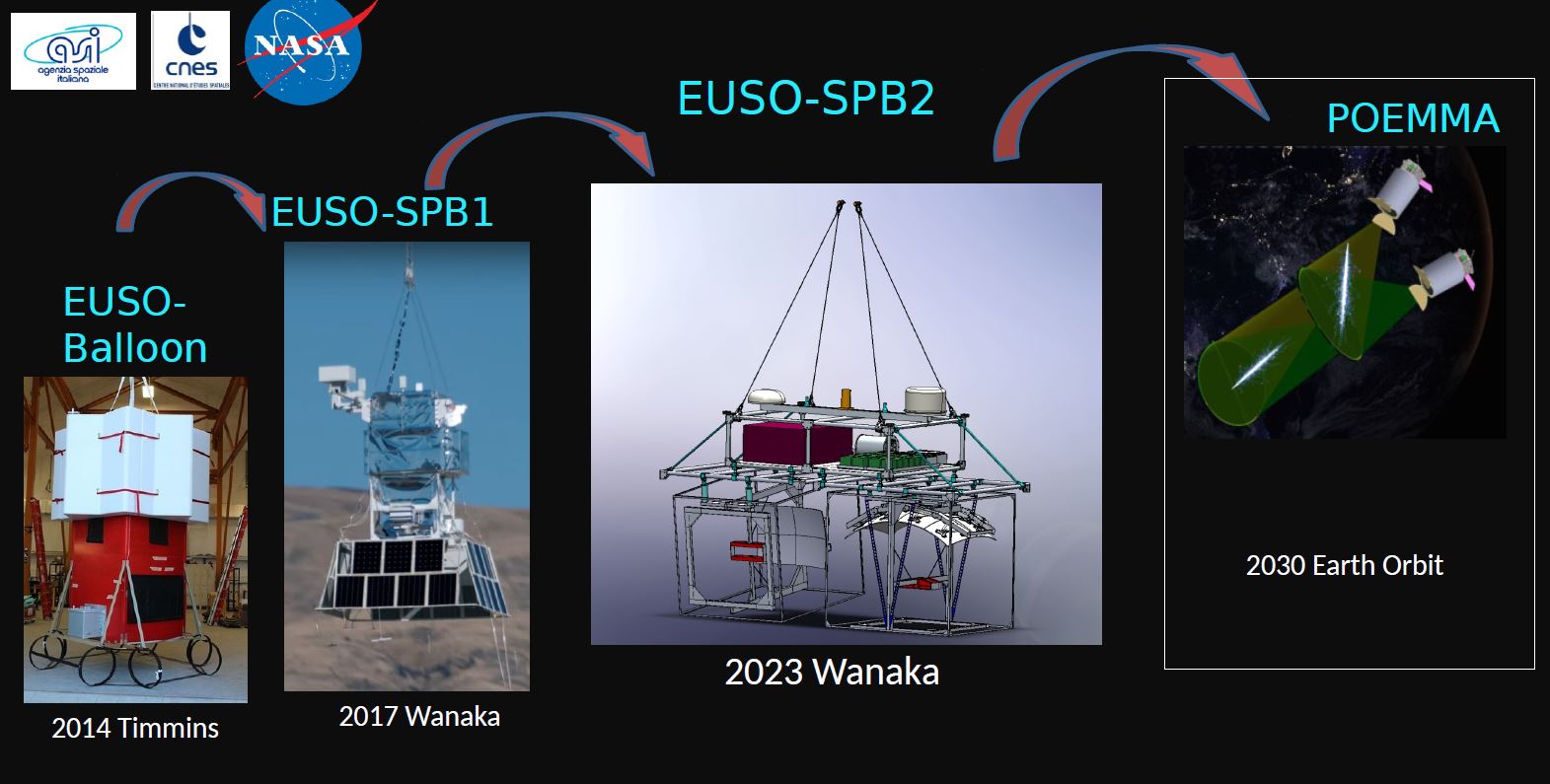}
    \caption{EUSO-SPB2 is a scientific and technological pathfinder for space based astroparticle experiments. It's missions objective is to measure UHECR with the fluorescence and direct \v{C}erenkov-technique from the stratosphere and search for high energy cosmic neutrinos. The design is based on the experience gained in precursors like EUSO-Balloon \cite{1ABDELLAOUI201954}, launched 2014 and EUSO-SPB2 \cite{Eser:2019M+}, launched 2017.}
    \label{fig:pathfinder}
\end{figure}
The Extreme Universe Space Observatory - Super Pressure Balloon (EUSO-SPB2) \cite{Wiencke:2019tX} is a cosmic ray detector, acting as a payload on board of a NASA super pressure balloon.
EUSO-SPB2 is primarily a technological and scientific pathfinder, see Fig. \ref{fig:pathfinder} and proof of concept for future space experiments, including POEMMA (Probe Of Multi-Messenger Astrophysics) \cite{Olinto_2021}.
 Going to space has the advantage that a large area of the sky can be covered and observed, to compensate the very small flux and rare occurrence of ultra high energy cosmic rays (UHECR).
EUSO-SPB2 will have the first observation of UHECRs from near-orbit altitude
with the fluorescence- and direct- \v{C}erenkov-technique at $33\ \text{km}$ or <7 mbar altitude. 
Below the limb it will search for tau neutrinos $\nu_\tau$ \cite{Cummings_2021}.
Target launch is April 2023 from Wanaka, New Zealand, with an anticipated flight duration of 100 days.

\section{EUSO-SPB2 detector}
\begin{table}[ht]
    \caption{The EUSO-SPB2 hardware specification for both telescopes.}
    \centering
\begin{tabular}{ |p{5cm}||p{4cm}|p{4cm}|  }
 \hline
 \multicolumn{3}{|c|}{EUSO-SPB2 specs} \\
 \hline
 Telescopes & Fluorescence FT & \v{C}erenkov CE \\
 \hline
Wavelength Sensitivity   & UV 300-420 nm    & no filter (300-900 nm)  \\
Energy Threshold &   EeV & PeV  \\
Sensor Type &MAPMT (Hamamatsu) & SiPM (S14521-6050CN) \\
Field of View & 3x(11x11)\degree  & (6.4x12.8)\degree \\
Pixel FOV& (0.2x0.2)\degree & (0.4x0.4)\degree  \\
Time frame & 1000 ns/bin & 10 ns \\
Pointing (zenith angle) &   nadir  & Earth's limb $\pm 10$\\
 \hline
\end{tabular}
    \label{tab:specs}
\end{table}
The hardware consists of two custom telescopes that measure \v{C}erenkov- and fluorescence-emission from extensive air-showers at the PeV and EeV-scale.
The fluorescence telescope camera consists of 3 Photo Detector Modules (PDMs) made of Multi Anode Photomultiplier Tubes (MAPMT, Hamamatsu) with a total of 6912 pixel. 
The instruments wide field of view (FoV) is $3\times \left( 11 \times 11 \right)  $\degree . 
A BG3 UV-transmitting band pass filter placed on top of the cameras focal surface gives a wavelength sensitivity of 300-420 nm. 
The time resolution is 1000 ns/GTU, where GTU is the gate time unit.
The fluorescence telescope will be in nadir position to look for the signature of extensive air showers (EAS) of UHECR, i.e. the fluorescence photon scattering from the electromagnetic component. \\
The \v{C}erenkov telescope camera consists of one camera made of Silicon Photomultiplier (SiPM, Hamamatsu (S14521-6050CN)) with 512 pixel. 
The instruments FoV is 6.4 \degree $\times$ 12.8\degree, so the primary mirror can be smaller in size. 
The wavelength sensitivity is UV-visible with 300-900 nm. 
The \v{C}erenkov telescope is a high speed detector with a time sensitivity of 10 ns.
It tilts $\pm 10$\degree\ to look towards the Earth's limb to see the emission of EAS produced by high energy $\nu_\tau$.
Both telescopes have an achromatic PMMA corrector plate at the 1-m diameter aperture, which is UV transparent. \cite{Marchi:2011qyc}. 
\section{Optics}
\begin{figure}[ht]
    \centering
\includegraphics[width=\textwidth]{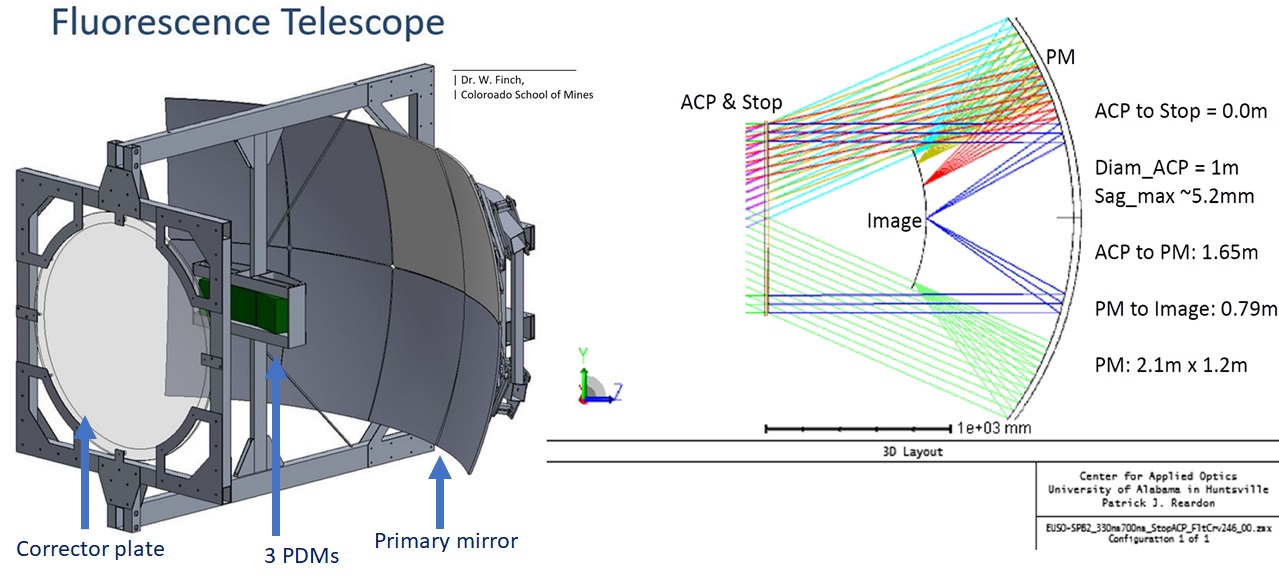}
    \caption{The fluorescence telescope design on the left and the corresponding raytracing model on the right. You can see the primary mirror PM, the achromatic corrector plate ACP with its diameter and maximum sag, as well as different distances between the components.}
    \label{fig:tel}
\end{figure}
\begin{figure}[ht]
    \centering
\includegraphics[width=\textwidth]{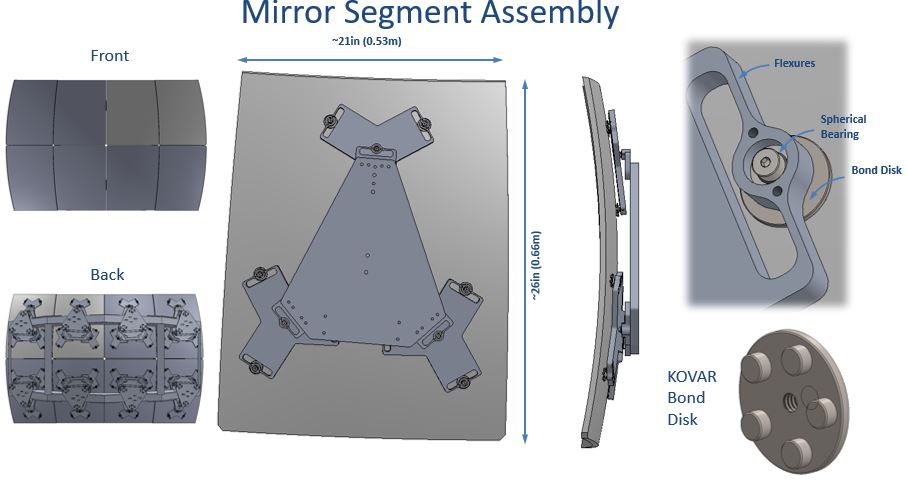}
    \caption{Mirror segment assembly with customized mirror segment attachment. The glass is epoxy glued to Kovar bonding disks. Kovar is a nickel–cobalt ferrous alloy and has approximately the same thermal expansion as glass.}
    \label{fig:mir}
\end{figure}

The telescopes feature modified Schmidt optics \cite{Schmidt}, with spherical borosilicate mirror glass.
The current design anticipates 6 mirror segments for the fluorescence and 4 mirror segments for the \v{C}erenkov telescope, where both have a radius of curvature ROC = 1659.8 mm with an effective focal length EFL = 860 mm.
For the \v{C}erenkov telescope a bi-focal mirror is used. 
One mirror in a row is tilted relative to the other one around vertical axis, with the optical axis is in horizontal position as seen in Fig. \ref{fig:tel}.
This creates a second separated spot at the focal surface, to distinguish between a charged particle which makes one flash on the camera and a light pulse from far-field outside the telescope, which makes two spots. 
The ratio of the collection area from the upper and lower mirrors depends on the direction of the incoming light, and affects the ratio of flux in the bifocal spots.
The fluorescence telescope is designed to measure track like signals.
Its 3 PDMs form a curved image surface, so that a field correcting or flattening lens is necessary to achieve a very small spot.
The light reflected by the primary mirror propagates through a plano-convex lens and a BG3 filter before it hits the elementary cell and the PMTS.
Without taking the BG3 filter into account, the overall throughput estimation, including corrector plate, mirror and filter losses is around 0.67. \\
On the back side the mirror segments of each telescope are assembled by a newly designed and fabricated holding structure as seen in Fig. \ref{fig:mir}.
Whippletrees hold around 20 lbs. mirror segments on nine attachment points that are glued to a bonding disk on spherical bearings. 
The bonding disks are made of Kovar, which has approximately the same thermal expansion as glass.
This design provides stability and high alignment precision, however, with an equally distributed force that orients the bond disks according the curvature of the mirror without bending it.
\\
\section{Testing}
\begin{figure}[ht]
    \centering
\includegraphics[width=\textwidth]{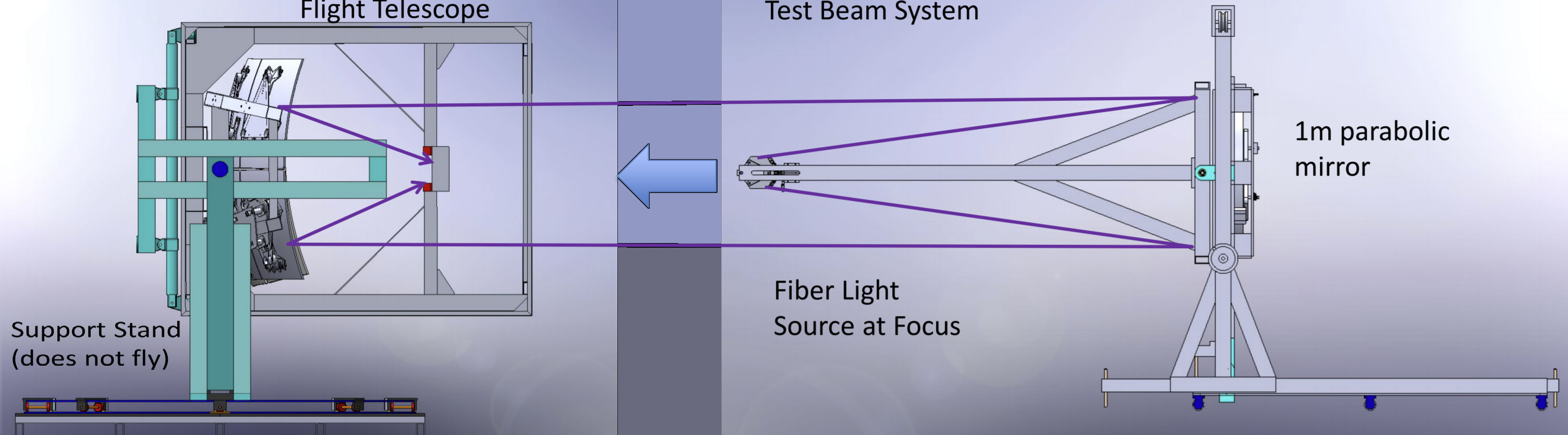}
    \caption{A schematic configuration of the fluorescence telescope on a supporting stand on the left and a 1-m diameter test beam system on the right. The test beam system is used to illuminate the telescope evenly with parallel light beam in the UV-light range. The main goal is to determine the PSF and efficiency. }
    \label{fig:1m}
\end{figure}

\begin{figure}[ht]
    \centering
\includegraphics[width=\textwidth]{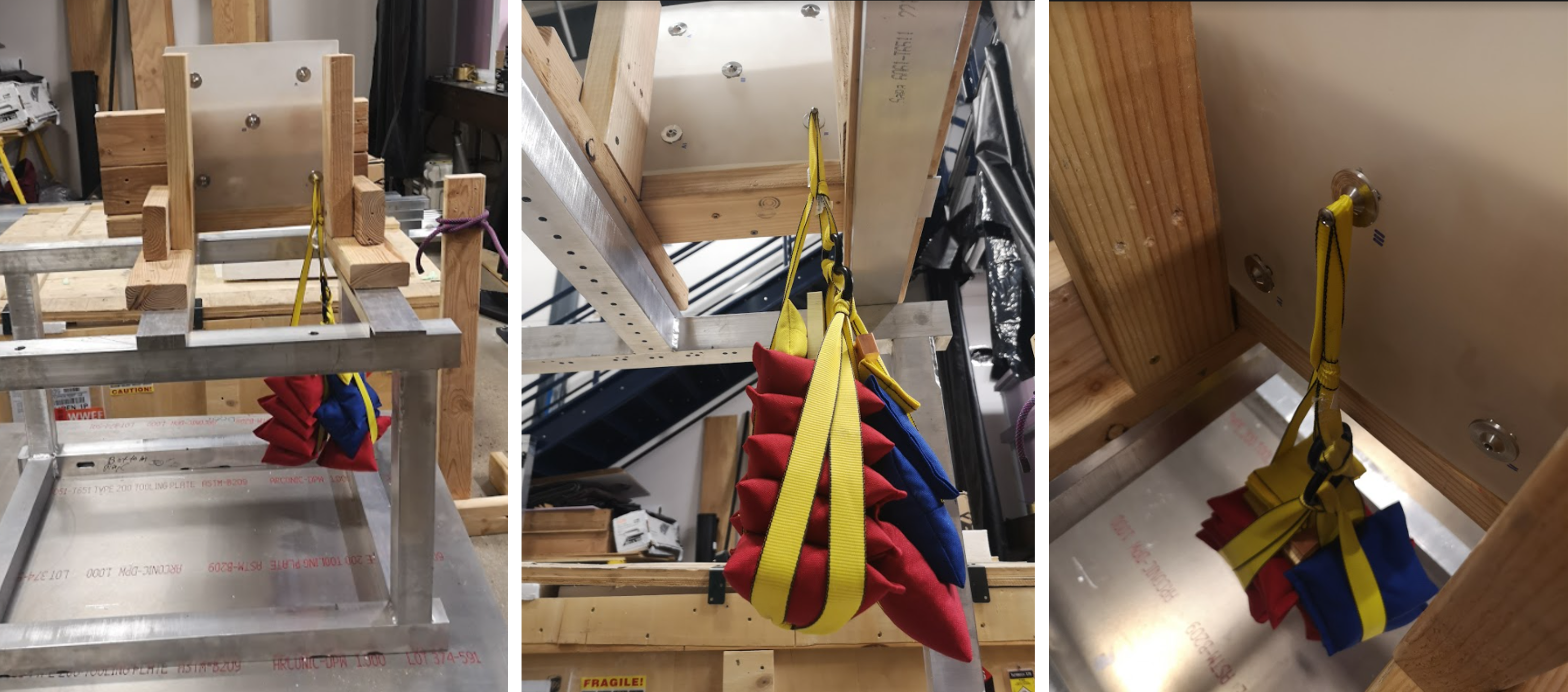}
    \caption{Test stand in shear or \v{C}erenkov configuration. The Kovar bond disks are epoxy glued to curved test glass piece, that has the same surface roughness as the primary mirror. The adhesive is under 43 lbs. shear stress with a small proportion of peeling stress in ambient. The test was successful and no break was recorded. Continuing test in various thermal and pressure conditions are currently under processing}
    \label{fig:shear}
\end{figure}

EUSO-SPB2 tests do verify design specifications that have to meet scientific and technological objectives.   
EUSO-SPB2 is a stratospheric experiment and differs mainly from ground based experiments in terms of its accessibility.
The design requirements are considerable more stringent, for example it has to meet thermal and mechanical requirements.
Additionally, flight hardware has redundancies, to reduce the risk of failure and ensure the missions success.
However, the main goals of the laboratory tests carried out at Colorado school of Mines are the determination of the point spread function (PSF) and efficiency of both telescopes, via a fabricated 1-m parallel test beam system, seen in Fig. \ref{fig:1m}.
\\
The idea is that LED emitted light for different UV-range wavelengths (340, 365, 385 nm) is reflected on a UV enhanced coated 1-m diameter parabolic F3.5 mirror, that provides a parallel beam illuminating the 1-m diameter aperture of the telescope.
The lights intensity will then be measured with the help of a diode, that will be placed directly at the aperture on a rotating stage, symmetrical as a windmill.
The stage will be operated remotely by stepper motors and a lead screw for a stable measurement where the step width and velocity is regulated.
Afterwards, the intensity will be measured in the area where around 95\% of the light accumulates, without the camera integrated.
For this a Cartesian robot system, again driven by stepper motors, will be used inside the mounting plates of the camera.
Finally, the camera is mounted and read out, with intensity and PSF being determined. \\
To make sure there is no displacement of the mirror segment, and thus a distortion of the spot size, the mirror alignment and its systems are tested in the lab.
There is no option to correct a shift or a degree of freedom in the mounting system while flight.
The mirror segments are epoxy glued to the metallic frame with EC-2216 adhesive, that is known for its high peel and shear strength, and durability over huge temperature cycles.
The glued system is also tested as part of the operational design requirements of NASA.
The glue joints are stress tested under a static uniformly distributed 20G load, in a thermal chamber for temperature below -40\degree C, and in a low pressure environment of <7 mbar, to simulate atmospheric conditions, see Fig. \ref{fig:shear} for an example of a glue joint being under 43 lbs of stress.
Before operating the telescopes the radius of curvature, with all mounted mirror segments will be confirmed.

\section{Field campaign}
\begin{figure}[ht]
    \centering
\includegraphics[width=\textwidth]{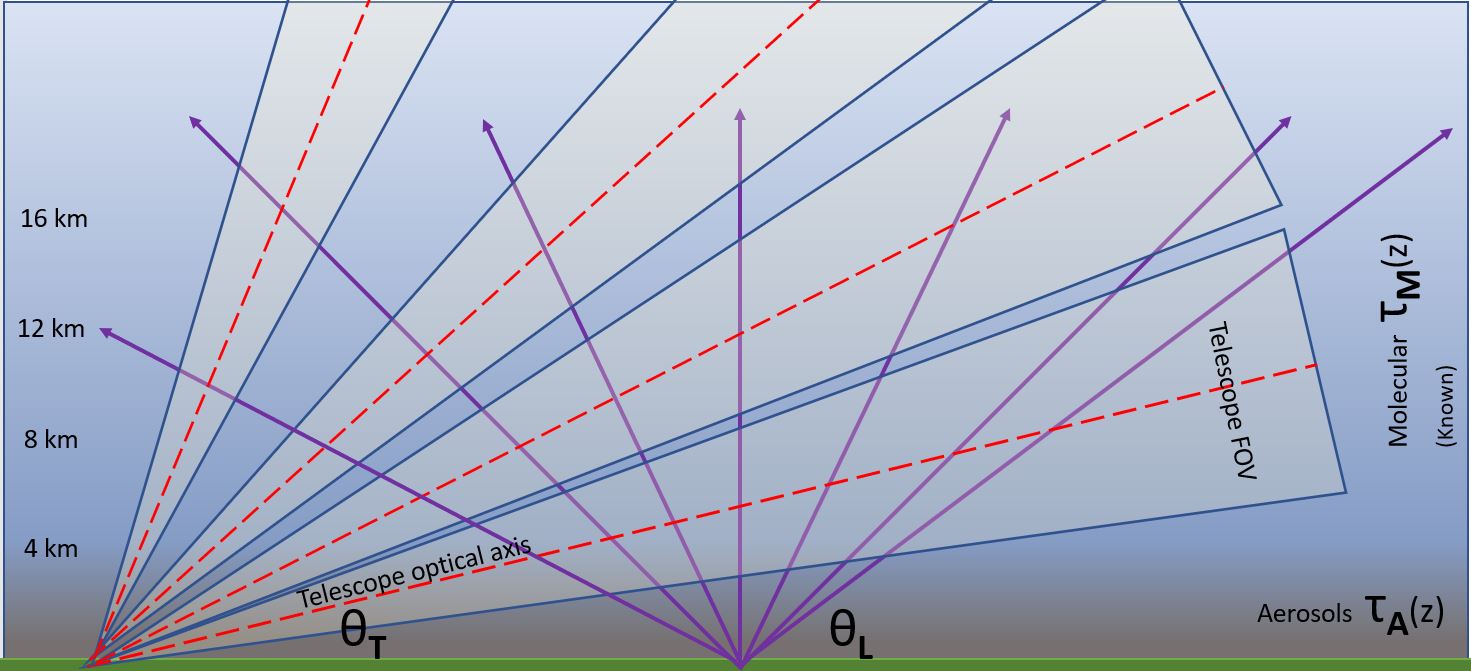}
    \caption{Sketch of the the field test configuration with the steerable telescope on the left and the steerable laser on the right with different inclination angles $\theta_T$ and $\theta_L$. The laser beam crosses the telescopes field of view high above the aerosol layer.}
    \label{fig:laser}
\end{figure}
Field tests will be performed in the desolate dark sites in the Western USA, in Utah, to check the detectors performance at moonless nights.
The main goal of the field campaign is to determine the field of view and the energy trigger threshold for both telescopes.
The campaign will include an absolute photometric calibration of the fully integrated instruments with the help of a bi-dynamic Lidar configuration, Fig \ref{fig:laser}. 
The laser system is inside a mobile utility trailer and can be leveled, so that the optical enclosure and the periscope is decoupled from the driving rack, when aiming in all directions above horizon.
It  will be possible to estimate the aerosol optical depth, without using the absolute calibration of the fluorescence telescope as a constraint, by steering the laser and the telescope to collect data over a range of distances and angles.
From this an absolute photometric calibration of the fluorescence telescope can be made knowing the laser energy.
This newly developed calibration method with a correction for aerosol attenuation will be tested first time.
Adopted Monte Carlo laser simulations have been made to predict the detectors performance, find a correlation and parametrization between laser energies and UHECR energies. \\

\textbf{Acknowledgment}: 
{\scriptsize This work was partially supported by Basic Science Interdisciplinary Research Projects of 
RIKEN and JSPS KAKENHI Grant (22340063, 23340081, and 24244042), by 
the Italian Ministry of Foreign Affairs	and International Cooperation, 
by the Italian Space Agency through the ASI INFN agreements n. 2017-8-H.0 and n. 2021-8-HH.0,
by NASA award 11-APRA-0058, 16-APROBES16-0023, 17-APRA17-0066, NNX17AJ82G, NNX13AH54G, 80NSSC18K0246, 80NSSC18K0473, 80NSSC19K0626, and 80NSSC18K0464 in the USA,  
by the French space agency CNES, 
by the Deutsches Zentrum f\"ur Luft- und Raumfahrt,
the Helmholtz Alliance for Astroparticle Physics funded by the Initiative and Networking Fund 
of the Helmholtz Association (Germany), 
by Slovak Academy of Sciences MVTS JEM-EUSO, by National Science Centre in Poland grants 2017/27/B/ST9/02162 and
2020/37/B/ST9/01821, 
by Deutsche Forschungsgemeinschaft (DFG, German Research Foundation) under Germany's Excellence Strategy - EXC-2094-390783311,
by Mexican funding agencies PAPIIT-UNAM, CONACyT and the Mexican Space Agency (AEM), 
as well as VEGA grant agency project 2/0132/17, and by by State Space Corporation ROSCOSMOS and the Interdisciplinary Scientific and Educational School of Moscow University "Fundamental and Applied Space Research".}
\bibliographystyle{unsrt}
\bibliography{skeleton.bib}

\clearpage
\section*{Full Authors List: \Coll\ Collaboration}


\scriptsize
\noindent
G.~Abdellaoui$^{ah}$, 
S.~Abe$^{fq}$, 
J.H.~Adams Jr.$^{pd}$, 
D.~Allard$^{cb}$, 
G.~Alonso$^{md}$, 
L.~Anchordoqui$^{pe}$,
A.~Anzalone$^{eh,ed}$, 
E.~Arnone$^{ek,el}$,
K.~Asano$^{fe}$,
R.~Attallah$^{ac}$
% ...
\\
\\
\noindent
% ...
$^{ac}$ LPR at Dept. of Physics, Faculty of Sciences, University Badji Mokhtar, Annaba, Algeria \\
% ...
$^{ah}$ Telecom Lab., Faculty of Technology, University Abou Bekr Belkaid, Tlemcen, Algeria \\
% ...
$^{cb}$ Universit\'e de Paris, CNRS, AstroParticule et Cosmologie, F-75013 Paris, France\\
% ...
$^{ed}$ Istituto Nazionale di Fisica Nucleare - Sezione di Catania, Italy\\
% ...
$^{eh}$ INAF - Istituto di Astrofisica Spaziale e Fisica Cosmica di Palermo, Italy\\
% ...
$^{ek}$ Istituto Nazionale di Fisica Nucleare - Sezione di Torino, Italy\\
$^{el}$ Dipartimento di Fisica, Universita' di Torino, Italy\\
% ...
$^{fe}$ Institute for Cosmic Ray Research, University of Tokyo, Kashiwa, Japan\\ 
% ...
$^{fq}$ Nihon University Chiyoda, Tokyo, Japan\\ 
% ...
$^{md}$ Universidad Polit\'ecnia de madrid (UPM), Madrid, Spain\\
% ...
$^{pd}$ University of Alabama in Huntsville, Huntsville, AL; USA\\
$^{pe}$ Lehman College, City University of New York (CUNY), NY, USA\\
% ...

\end{document}